\documentclass[12pt, a4paper]{article}

\usepackage[margin=2.5cm]{geometry}
\usepackage{helvet,times}
\usepackage{bm,textcomp}
\usepackage{amssymb}
\usepackage{amsmath}
\usepackage{graphicx}
\usepackage{subfigure}
\usepackage[percent]{overpic}
\usepackage[table]{xcolor}
\usepackage{marvosym}
\usepackage{appendix}
\usepackage{rotating}
\usepackage[round,sort&compress]{natbib}

\newcommand{\envelope}{(\raisebox{-.5pt}{\scalebox{1.45}{\Letter}}\kern+0.1pt)}
\newcommand{\brac}[1]{\left( {#1} \right)}    
\newcommand{\bracs}[1]{\left[ {#1} \right]}   

\begin{document}
\title{A computational study of stress fiber-focal adhesion dynamics governing cell contractility}
\author{
M. Maraldi$^1$, C. Valero$^2$, K. Garikipati$^{1, 3}$\footnote{correspondence: krishna@umich.edu} \\
\small{$^1$Department of Mechanical Engineering, University of Michigan, Ann Arbor, Michigan} \\ 
\small{$^2$M2BE, Arag\'{o}n Institute of Engineering Research (I3A), University of Zaragoza, Zaragoza, Spain} \\ 
\small{$^3$Department of Mathematics, University of Michigan, Ann Arbor, Michigan}
}
\date{}
\maketitle
\begin{abstract}
{
We apply a recently developed model of cytoskeletal force generation to study a cell's intrinsic contractility, as well as its response to external loading.
The model is based on a non-equilibrium thermodynamic treatment of the mechano-chemistry governing force in the stress fiber-focal adhesion system.
Our computational study suggests that the mechanical coupling between the stress fibers and focal adhesions leads to a complex, dynamic, mechano-chemical response.
We collect the results in response maps whose regimes are distinguished by the initial geometry of the stress fiber-focal adhesion system, and by the external load on the cell.
The results from our model connect qualitatively with recent studies on the force response of smooth muscle cells on arrays of polymeric microposts (Mann et al., \textit{Lab. on a Chip}, \textbf{12}, 731-740, 2012).
}
\end{abstract}
%
\section*{INTRODUCTION}
In contractile cells, such as smooth muscle cells and fibroblasts, the generation of traction force is the result of two different actions: myosin-powered cytoskeletal contractility and external mechanical stimuli (applied stretch or force).
The cooperation between these two aspects determines the level of the force within the cell and influences the development of cytoskeletal components via the (un)binding of proteins.
Important cytoskeletal components that mediate this interplay of mechanics and chemistry are stress fibers and focal adhesions.

Stress fibers are bundles of 10--30 actin filaments held together by the binding protein $\alpha$-actinin \citep{pellegrin:2007}; fascin, epsin, filamin and myosin, among other proteins, have also been detected in stress fibers.
Cytoskeletal contractility originates from the action of myosin molecules, which attach themselves to the actin filaments and step along them, causing anti-parallel filaments to slide past each other, thus generating a contraction of the stress fiber.
The speed at which filament slide past each other decreases with tensile force \citep{hill:1938}.
The myosin stepping rate reaches a stall at some critical value of tensile force, at which contractility ceases.

The binding rates of actin and myosin (and presumably of other proteins, also) into the stress fiber is force-dependent \citep{pollard:2003}; within some regime of tensile force \textit{auto}-generated by stress fiber contractility, the binding rates appear to be boosted, and the fibers grow in thickness \citep{chrzanowska:1996, ingber:2003}.
Eventually, a sufficiently high force, perhaps externally applied, must cause rapid unbinding of the proteins and cytoskeletal disassembly.
The complexity of this mechano-chemical response is enhanced because the stress fibers also demonstrate, besides the aforementioned active response due to myosin action, a passive viscoelastic force-stretch behavior \citep{kumar:2006}.

Focal adhesions are integrin-containing transmembrane structures that anchor the cytoskeletal stress fibers to the extra-cellular matrix (ECM).
In addition to integrin they contain scores of other proteins including paxillin, tensin, focal adhesion kinase, talin and vinculin.
The latter two proteins connect the integrins to f-actin in the stress fibers, to complete the linkage of the cytoskeleton to the ECM.
However, focal adhesions are not merely static anchors.
They themselves demonstrate a complex dynamics of growth, disassembly, and even a translational mode in which they appear to slide over the interface between the cell membrane and ECM, strikingly shown by Nicolas and co-workers \citep{nicolasetal:2004}.
These regimes of the dynamics are caused by (un)binding of focal adhesion proteins, and notably are force-sensitive; cytoskeletal contractility forces as well as externally applied loads may elicit this mechanosensitive response \citep{riveline:2001, balaban:2001}.

It is inevitable that the combination of two such mechano-chemically dependent systems (stress fibers and focal adhesions) in the cytoskeleton leads to a rich dynamic response, where the force as well as the systems' structures themselves continuously evolve.
Some of these aspects have been addressed in the literature, and a variety of models have been proposed which study stress fibers and focal adhesions separately \citep{besser:2011, besser:2007, kaunas:2010, kaunas:2008, kruse:2000, stachowiak:2008, stachowiak:2009} or, in some cases, in combination \citep{walcott:2010, harland:2011, deshpande:2006, deshpande:2008}.
They focus on different aspects of the problem, such as cell traction \citep{tan:2003}, effects of substrate stiffness \citep{engler:2006, chan:2008}, cell shape \citep{chen:2003}, cell contractility \citep{peterson:2004}, cytoskeletal orientation under dynamic load \citep{franke:1984, kaunas:2005, kaunas:2008, wei:2008}, and stress fiber viscoelasticity \citep{kumar:2006, peterson:2004}.
Some studies also address the role of the small GTPases, Rho and Rac, in regulating stress fiber formation \citep{ridley:1992a, ridley:1992b, sander:1999}.

In this paper, we use a recently developed model for the coupled mechano-chemical response of stress fiber-focal adhesion systems to study the development of contractile force, as well as the behavior of such systems under load.
The model is based on non-equilibrium thermodynamics and has been described in detail in a work by Maraldi and Garikipati \citep{maraldi:2013}.
Our focus here is on the modes of generation and decay of the force in the system, as well as on the growth and disassembly of the stress fibers and focal adhesions.
These questions are addressed in the context of both cell contractility in absence of external load and system response to an external stretch.
Our motivation comes from studies of force generated by smooth muscle cells plated on micropost arrays \citep{mann:2012}; nevertheless, our model is capable of much greater detail than is accessible experimentally.
The experiments demonstrate variability in the response both between cells and between individual stress fibers in the same cell; accordingly, our aim is to reproduce the broad trends seen in the experiments and provide a key to interpret the response variability observed in the experiments, while examining in greater detail the underlying mechano-chemical dynamics that the model reveals.
%
\section*{THE UNDERLYING MODEL}
\label{sect:model}
\begin{figure}[h]
\centering
\includegraphics{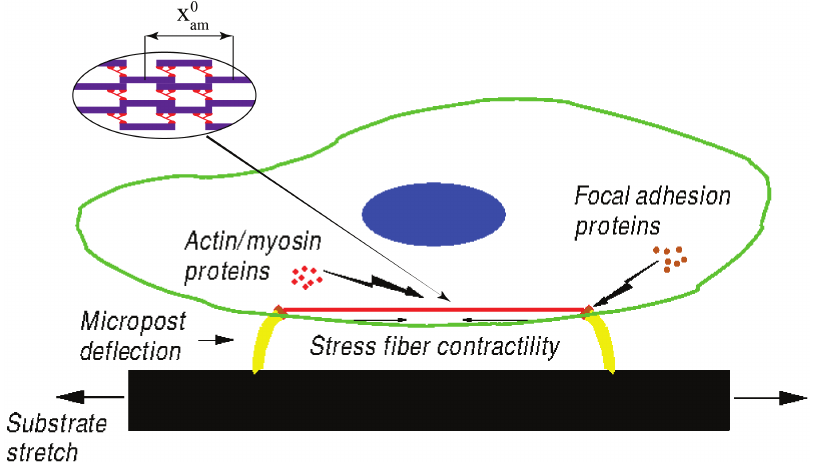}
\caption{\small{Schematic representation of the model.}}
\label{fig:model}
\end{figure}

Calculations were carried out using a modified version of a model proposed by Maraldi and Garikipati \citep{maraldi:2013}; the model does not include chemical signaling in order to explicitly highlight the role of mechanical force as a signal, instead.
The original layout has been adapted to include the presence of elastic microposts, in order to simulate the behavior of the stress fiber - focal adhesion ensemble under the conditions of the experimental tests performed by Mann et al. \citep{mann:2012}.
Specifically, the model adopted here (Fig. \ref{fig:model}) consists of a stress fiber connected to a focal adhesion at each end, with each focal adhesion being attached to the top of a PDMS micropost, and an elastic (PDMS) substrate underlying the microposts; the cytosolic reservoir supplying proteins to the stress fiber and focal adhesions is also included.
The substrate can be stretched to introduce an external mechanical loading of the system.
The stress fiber and the focal adhesions are mechano-chemical subsystems formed by assembly of representative proteins supplied by the cytosol.

To substantiate the discussion of the results provided in this paper, we briefly report the main concepts behind the model we adopted; further details can be found in Appendix and elsewhere \citep{olberding:2010,maraldi:2013}.
In the model, the stress fiber is considered to span between two focal adhesions and to be always under tension; hence, its reference length, $x_\mathrm{sf}^0$, is fixed, as it would not be possible to add proteins at its ends without first abrogating the tension.
Protein binding/unbinding therefore only affects the thickness of the stress fiber.
On the other hand, protein binding/unbinding is allowed to occur anywhere along the focal adhesion; however, to compute the relevant kinematic quantities for this sub-system, only the binding rates at its ends need to be tracked.
The force is assumed to be uniformly distributed along the focal adhesion and through the stress fiber's thickness \citep{maraldi:2013}.

The number of stress fiber representative proteins and the focal adhesion distal and proximal ends' positions ($N_\mathrm{sf}$, $x_\mathrm{d}$ and $x_\mathrm{p}$, respectively) are the variables tracked with respect to time.
The ordinary differential equations constituting the model are \citep{olberding:2010,maraldi:2013}:
\begin{equation} \label{eq:sf-evo}
\dot{N}_\mathrm{sf} = \! \!
  \begin{cases}
    k_\mathrm{sf}^\mathrm{b} \brac{N^\mathrm{max}_\mathrm{sf} \!-\! N_\mathrm{sf}} \brac{ 1-e^{\brac{\mu_\mathrm{sf}-\mu^\mathrm{sf}_\mathrm{cyt}}/k_B T} }, & \mkern-18mu \text{ \textit{(b)}}\\
    k_\mathrm{sf}^\mathrm{u} \ e^{\chi_\mathrm{sf}} \brac{ e^{-\brac{\mu_\mathrm{sf}-\mu^\mathrm{sf}_\mathrm{cyt}}/k_B T} -1 }, & \mkern-18mu \text{ \textit{(u)}}
  \end{cases}
\end{equation}
\begin{equation} \label{eq:fa-evo-d}
\dot{x}_\mathrm{d} = - \lambda^2 \!
  \begin{cases}
    k_\mathrm{fa}^\mathrm{b} \brac{ 1-e^{\brac{\mu^\mathrm{d}_\mathrm{fa}-\mu^\mathrm{fa}_\mathrm{cyt}}/k_B T} }, & \text{ \textit{(b)} }\\
    k_\mathrm{fa}^\mathrm{u} \ e^{\chi_\mathrm{fa}} \brac{ e^{-\brac{\mu^\mathrm{d}_\mathrm{fa}-\mu^\mathrm{fa}_\mathrm{cyt}}/k_B T} -1 }, & \text{ \textit{(u)}, }
  \end{cases}
\end{equation}
\begin{equation} \label{eq:fa-evo-p}
\dot{x}_\mathrm{p} =   \lambda^2 \!
  \begin{cases}
    k_\mathrm{fa}^\mathrm{b} \brac{ 1-e^{\brac{\mu^\mathrm{p}_\mathrm{fa}-\mu^\mathrm{fa}_\mathrm{cyt}}/k_B T} }, & \text{ \textit{(b)} }\\
    k_\mathrm{fa}^\mathrm{u} \ e^{\chi_\mathrm{fa}} \brac{ e^{-\brac{\mu^\mathrm{p}_\mathrm{fa}-\mu^\mathrm{fa}_\mathrm{cyt}}/k_B T} -1 }, & \text{ \textit{(u)}, }
  \end{cases}
\end{equation}
where the label \textit{(b)} indicates the equations used for the case in which $\mu_\mathrm{\alpha}-\mu^\mathrm{\alpha}_\mathrm{cyt} \leq 0$ (proteins binding) for sub-system $\alpha = \mathrm{sf}, \mathrm{fa}$, whereas \textit{(u)} indicates the equations for the case in which $\mu_\mathrm{\alpha}-\mu^\mathrm{\alpha}_\mathrm{cyt} \geq 0$ (proteins unbinding).
In Eq. (\ref{eq:sf-evo}), $\mu_\mathrm{sf}$ is the chemical potential of representative proteins in the stress fiber, $\mu^\mathrm{sf}_\mathrm{cyt}$ is the chemical potential of stress fiber proteins in the cytosolic reservoir, and $N_\mathrm{sf}^\mathrm{max}$ is the maximum number of stress fiber proteins available to the given stress fiber.
In Eq. (\ref{eq:fa-evo-d}) and Eq. (\ref{eq:fa-evo-p}), $\mu^\mathrm{d}_\mathrm{fa}$ and $\mu^\mathrm{p}_\mathrm{fa}$ are the chemical potentials of the proteins in the focal adhesion evaluated at its distal and proximal ends, respectively, $\mu_\mathrm{cyt}^\mathrm{fa}$ is their chemical potential in the cytosol and $\lambda$ is the size of a focal adhesion complex.
For the detailed expressions of the chemical potentials see Eqs.\eqref{eq:chem-pots} in Appendix .
Moreover, $k^\mathrm{b}_\alpha$, $k^\mathrm{u}_\alpha > 0$ are respectively the binding and unbinding coefficients for sub-system $\alpha$, $k_B$ is the Boltzmann constant, and $\chi_{\alpha} = \chi_{\alpha} \brac{P}$ is a force-dependent exponent regulating the rapid dissociation of molecular bonds \citep{bell:1978, maraldi:2013}.
We note that the form of Eqs. (\ref{eq:sf-evo}--\ref{eq:fa-evo-p}) comes from classical non-equilibrium thermodynamics, and incorporates the assumption of local equilibrium \citep{degrootmazur:1984}.

Mechanical equilibrium is assumed to hold; hence, the forces developed within the stress fiber, the focal adhesions and the microposts are equal to one another and identified as the force within the system: $P = P_\mathrm{sf} = P_\mathrm{fa} = P_\mathrm{mp}$.
The determination of $P$ is essential for calculating the chemical potentials of the focal adhesion, the stress fiber and the cytosol, which are the driving forces for the chemical processes \citep{maraldi:2013} and appear in the rate equations Eqs.(\ref{eq:sf-evo}--\ref{eq:fa-evo-p}).

In the Discussion Section, we will observe that the stress fiber's constitutive nature plays a major role in the complex mechanical response of the system.
Indeed, the contractile and viscoelastic features of the stress fiber strongly influence the development of the force within the whole system.
In particular, the force developed within the stress fiber (and consequently within the whole system, due to mechanical equilibrium) can be expressed as the sum of three different contributions: $P_\mathrm{sf} = P_\mathrm{sf}^\mathrm{e} + P_\mathrm{sf}^\mathrm{ve} + P_\mathrm{sf}^\mathrm{ac}$, where $P_\mathrm{sf}^\mathrm{e}$ is the elastic component, $P_\mathrm{sf}^\mathrm{ve}$ accounts for the viscous response and $P_\mathrm{sf}^\mathrm{ac}$ is the active contractile force.
Fig. \ref{fig:model} also shows the actomyosin contractile units that make up the stress fiber.
Each unit consists of one myosin motor and one half-length of each interleaved, anti-parallel actin filament that the motor causes to intercalate.
The units also are assumed to have the same length, and the total number of contractile units is therefore proportional to $N_\mathrm{sf}$.
We take each such unit to have the same strain rate in the stress fiber.
See Eqs. (\ref{eq:forces}) and the ensuing discussion in Appendix for the complete active contractile force model.

A specific set of parameters was chosen (Tab. \ref{tab:params} in Appendix) and the model was tested for its ability to reproduce the main features of the force response of smooth muscle cells plated on an array of polymeric microposts \citep{mann:2012}.
To access a variety of responses, the initial stress fiber length was varied over a range typically reported for a cell ($10$--$65\; \mu\mathrm{m}$), while the initial focal adhesion length was varied in the $0$--$2\; \mu\mathrm{m}$ range.
For the tests in which an external load was applied to the system, the extent of the substrate stretch was varied between $0.05$ and $0.15$, to make connections with Mann et al. \citep{mann:2012}.

\section*{RESULTS}
\label{sect:res}
%
\begin{figure}[h]
\centering
\includegraphics{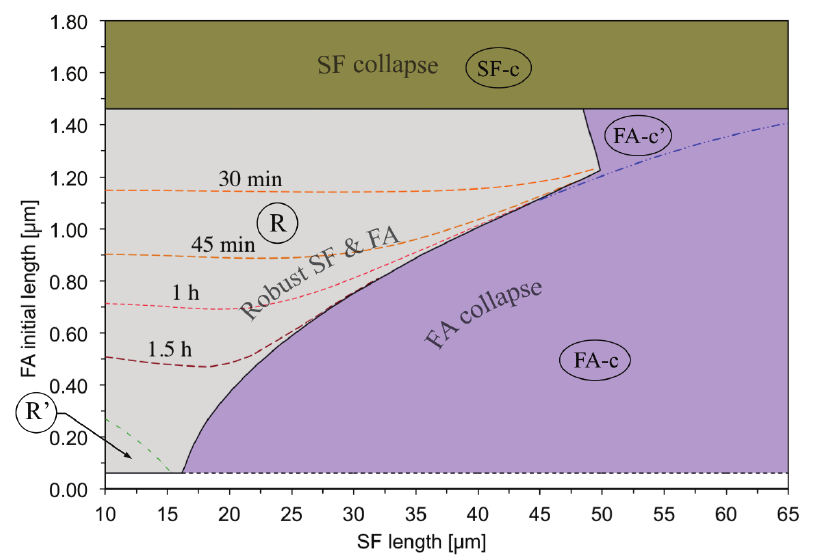}
\caption{\small{System response map with no applied stretch. \textit{R} is the region in which the stress fiber and the focal adhesion reach full development (\textit{robust stress fiber and focal adhesion} region). The dashed curves are iso-time contours of micropost coverage by the growing focal adhesion. \textit{R}$^\prime$ is the region in which focal adhesion translation causes stress fiber force relaxation to zero. The system collapses in regions \textit{FA-c} and \textit{FA-c}$^\prime$ due to focal adhesion resorption (\textit{focal adhesion collapse} regions), and in \textit{SF-c} due to stress fiber resorption (\textit{stress fiber collapse} region).}}
\label{fig:phase-d-un}
\end{figure}
\section*{System response map and collapse mechanisms with no applied strain}
\label{sect:maps-unstretchd}
We first seek to model the contractility of a cell on an array of microposts when the substrate is not subjected to an external strain.
The corresponding system responses are collected in the map of Fig. \ref{fig:phase-d-un}.

In region \textit{R}, a robust, fully developed system is obtained, with a stable stress fiber and a growing focal adhesion.
Fig. \ref{fig:phase-d-un} shows that this region may extend down to $\hat{x}_\mathrm{fa}^0 = \lambda = 58$ nm (black dashed line in Fig. \ref{fig:phase-d-un}), which is the size of a single complex of focal adhesion proteins, and represents the smallest focal adhesion in our model.\footnote{The term \textit{focal complex} may be more appropriate in this limit.}
Notably, even this smallest initial focal adhesion gives rise to a robust system if $x_\mathrm{sf}^0$ is small. Region \textit{R} spans a wider range of $\hat{x}_\mathrm{fa}^0$ values than any other region.
However, for larger values of $x_\mathrm{sf}^0$ this range of $\hat{x}_\mathrm{fa}^0$ becomes increasingly narrow, as other failure mechanisms become dominant (regions \textit{FA-c} and \textit{SF-c}).

Inside region \textit{R} in Fig. \ref{fig:phase-d-un}, the system exhibits different behaviors, some of which are induced by the fact that the focal adhesion is constrained to develop on the surface of the micropost, which has finite area.
The dashed curves indicate the times at which the focal adhesion has grown to the size of the micropost diameter.
Smaller $\hat{x}_\mathrm{fa}^0$ translates to greater growth times, as would be expected.
Further details are provided in the following sub-section.
The dash-dot green line in Fig. \ref{fig:phase-d-un} delimits the sub-region \textit{R}$^\prime$, characterized by low values of $x_\mathrm{sf}^0$ and $\hat{x}_\mathrm{fa}^0$.
For these configurations the system does not collapse, but the stress fiber force vanishes at small times.
Here, treadmilling of proteins through the cytosol allows the focal adhesion structure to translate in the direction of the force, causing the force in the system to relax to zero (blue curves in Fig. \ref{fig:region-A-evos}).

Outside region \textit{R}, the system collapses due to different failure mechanisms; in region \textit{FA-c}, characterized by low values of the $\hat{x}_\mathrm{fa}^0 / x_\mathrm{sf}^0$ ratio, the collapse is due to the complete resorption of the focal adhesion.
Under these conditions, in fact, the stress fiber is able to generate a high active force, $P^\mathrm{ac}_\mathrm{sf}$; as a consequence, the force within the system $P$ is high and exceeds the focal adhesion's ability to sustain mechanical load (see the discussion on the focal adhesion critical load in the Discussion Section), causing its complete resorption by unbinding at its distal end.
Similarly, for high values of $x_\mathrm{sf}^0$ (subregion \textit{FA-c}$^\prime$) the system experiences focal adhesion collapse due to the finite surface area of the micropost which constrains the growth of the focal adhesion (See Appendix \ref{sect:evolutions-unstretched-FAc} for details).

In region \textit{SF-c} the system collapses due to stress fiber failure.
The large focal adhesion increases the system stiffness so that a high force $P$ can be developed under strain control. This ultimately causes stress fiber resorption, and the system collapses even as the large focal adhesion survives.

\section*{Time-dependent response of the system with no applied strain}
\label{sect:evolutions-unstretched}
\begin{figure*}[h]
\centering
\includegraphics{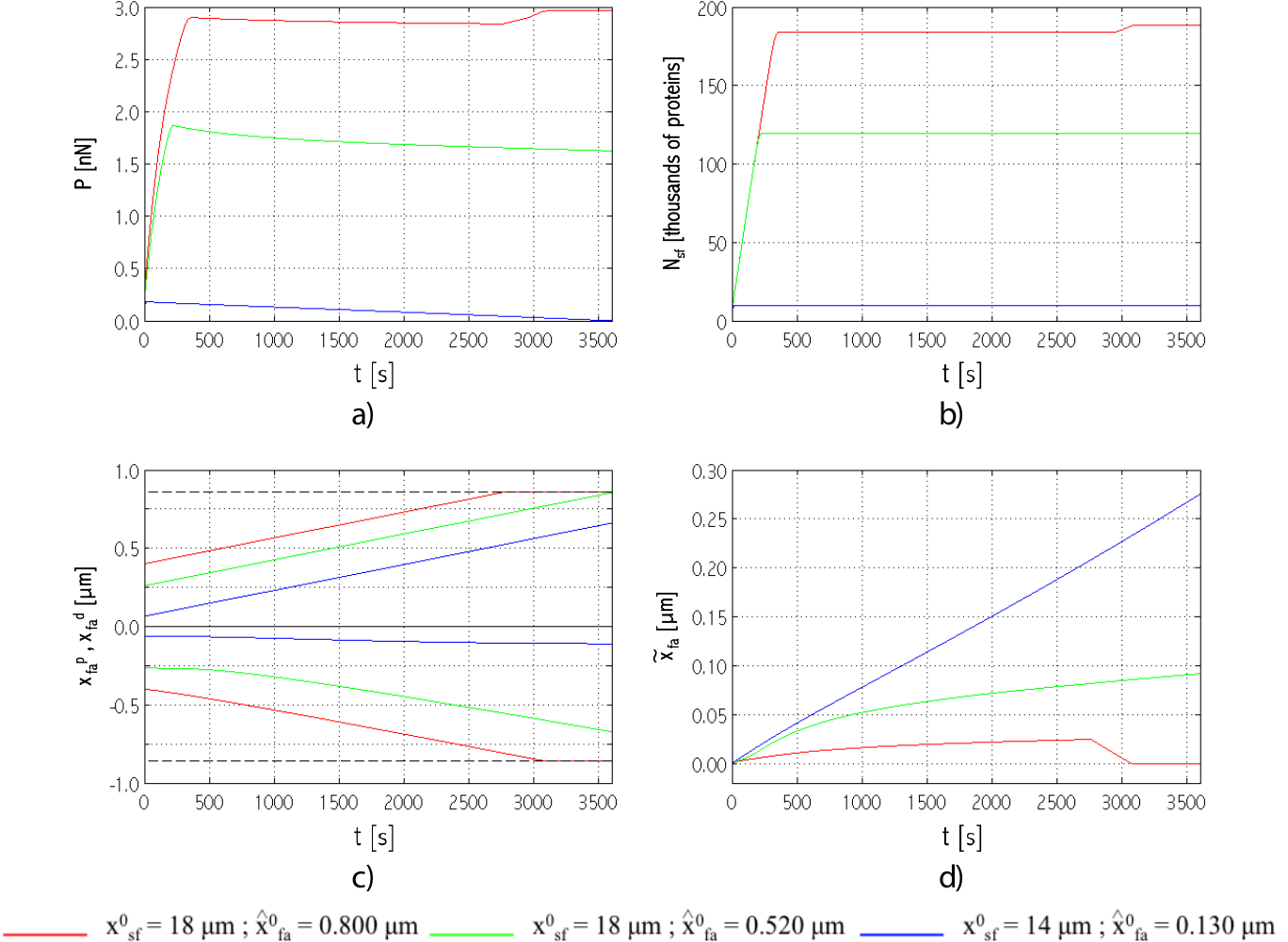}
\caption{\small{Time evolution of a) force, $P$; b) number of actin monomers in the stress fiber, $N_\mathrm{sf}$; c) focal adhesion distal (negative values, $x^\mathrm{d}_\mathrm{fa}$) and proximal (positive values, $x^\mathrm{p}_\mathrm{fa}$) ends positions; and d) focal adhesion centroid position, $\tilde{x}_\mathrm{fa}$ for three different system initial configurations belonging to region \textit{R} in Fig. \ref{fig:phase-d-un}. In c) the dashed black lines indicate the position of the micropost edges.}}
\label{fig:region-A-evos}
\end{figure*}

The detailed dynamics of the system in terms of the time evolution of force $P$, number of proteins in the stress fiber $N_\mathrm{sf}$, position of the focal adhesion proximal and distal ends ($x_\mathrm{fa}^\mathrm{p}$ and $x_\mathrm{fa}^\mathrm{d}$, respectively) and centroid position $\tilde{x}_\mathrm{fa}$ are depicted in Fig. \ref{fig:region-A-evos} for three typical system configurations belonging to region \textit{R} of the response map in Fig. \ref{fig:phase-d-un}.

For configurations in region \textit{FA-c} of the response map, a similar discussion is provided in Appendix.
The force within the system, $P$, is often referred to as the \textit{contractile force} in literature; however, we prefer not to use this terminology, as, according to the stress fiber constitutive model used for the present study \citep{maraldi:2013}, this force depends not only on contractility, but also on the passive elastic or viscoelastic response of all the sub-systems (see Eqs. (\ref{eq:forces}) in Appendix, and the discussion on stress fiber rheology in the Model Section) and on loads external to the system (see the discussion related to Fig. \ref{fig:stretched-evos}).

\begin{figure}[t]
\centering
\includegraphics{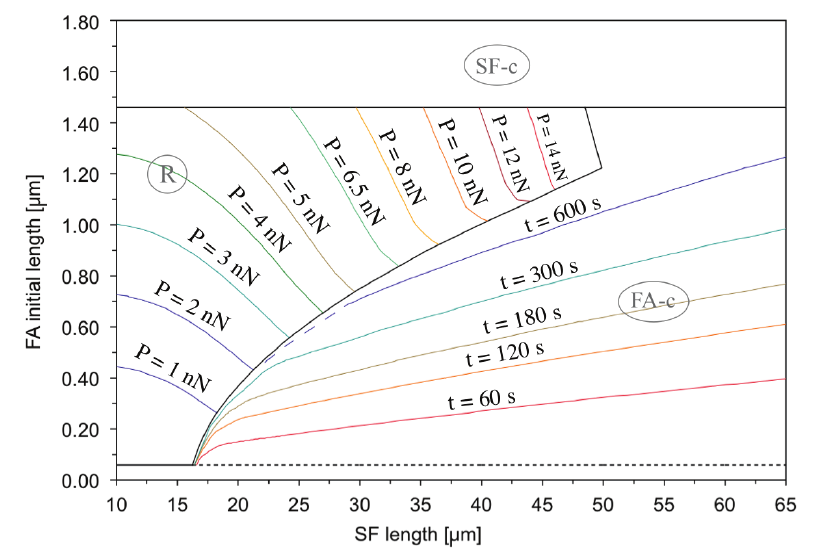}
\caption{\small{Contour plots of the maximum contractile force $P$ and of the focal adhesion resorption time $t$ for configurations belonging, respectively, to regions \textit{R} and \textit{FA-c} of the map in Fig. \ref{fig:phase-d-un}.}}
\label{fig:contours-unstretched}
\end{figure}
Fig. \ref{fig:region-A-evos}a shows the evolution of $P$; in all cases, the force initially increases and, after a time interval that depends on the initial values $\hat{x}_\mathrm{fa}^0$ and $x_\mathrm{sf}^0$, it attains a near-plateau characterized by a negative slope.
Accordingly, $N_\mathrm{sf}$ increases until a critical concentration is reached at which protein recruitment stops (Fig. \ref{fig:region-A-evos}b).

Although the cases in Fig. \ref{fig:region-A-evos} all fall into region \textit{R}, the detailed dynamics differ notably from one another.
The blue curves, for instance, refer to a configuration in region \textit{R}$^\prime$: while $P$ completely vanishes, neither the stress fiber nor the focal adhesion dissolve, as shown, respectively, in Figs. \ref{fig:region-A-evos}b and \ref{fig:region-A-evos}c.
Indeed, the stress fiber grows continuously until the aforementioned critical actin concentration is attained and the focal adhesion also grows, by addition of complexes at both its ends.
Interestingly, the relaxation of $P$ corresponds with focal adhesion translation due to protein treadmilling, as seen in the evolving position of the focal adhesion centroid (Fig. \ref{fig:region-A-evos}d).
Region \textit{R}$^\prime$ may therefore be regarded as an enhanced translation region.

The red curves in Fig. \ref{fig:region-A-evos} show the system dynamics when the finite cross-section of the micropost limits focal adhesion growth.
A \textit{stiffening} effect is imposed on the system (as seen from the red curve in Fig. \ref{fig:region-A-evos}a, at $t \simeq 3000$ s).
As shown in Fig. \ref{fig:region-A-evos}c, the faster-growing proximal end of the focal adhesion is first to reach the corresponding micropost edge (this holds for all system configurations).
Consequently, the focal adhesion continues to grow only at the distal end, and its centroid, which is the center of action of the stress fiber force, moves backward (Fig. \ref{fig:region-A-evos}d).
The focal adhesion translation away from the direction of the force induces a kinematic stiffening - in the same manner as a translation in the direction of the force induces a kinematic relaxation (see the preceding discussion, as well as the forthcoming one on the competition between stress fiber contractility and focal adhesion translation) - which makes the chemical potential term $(\mu_\mathrm{sf} - \mu^\mathrm{sf}_\mathrm{cyt})$ of Eq. \ref{eq:sf-evo} negative and re-establishes a growth regime for the stress fiber.
Consequently, more actin and myosin are recruited to the stress fiber and $P$ starts rising again until the slower-growing distal end of the focal adhesion reaches the corresponding micropost edge.
The focal adhesion has no more room for growth, $N_\mathrm{sf}$ reaches a second, higher, critical concentration and the contractile force plateaus out.
The stress fiber-focal adhesion system is at equilibrium in this case.

When the system configuration falls outside region \textit{R}$^\prime$ of the response map in Fig. \ref{fig:phase-d-un} and neither end of the focal adhesion reaches the micropost edge, the dynamics follow the green curves of Fig. \ref{fig:region-A-evos}: the critical value of $N_\mathrm{sf}$ is reached in the stress fiber, which stops growing, while the focal adhesion continues to grow by recruiting complexes at both ends (Fig. \ref{fig:region-A-evos}c).
The observed force relaxation is related to translation as explained above.

The maximum value attained by the force in the system, $P$, is of interest for robust systems; it depends on $x_\mathrm{sf}^0$ and $\hat{x}_\mathrm{fa}^0$, as reported in the contour plot of Fig. \ref{fig:contours-unstretched} for configurations in region \textit{R}.
It can be noted that a higher $x_\mathrm{sf}^0$ results in a higher value of the maximum of $P$. However, $\hat{x}_\mathrm{fa}^0$ also has some influence: especially for low $x_\mathrm{sf}^0$, a high $\hat{x}_\mathrm{fa}^0$ leads to an increased maximum $P$.
Turning to the stress fiber growth, the maximum, or critical, value of $N_\mathrm{sf}$ is proportional to the stress fiber radius $r_\mathrm{sf}$ and to the number of actin filaments $N_\mathrm{fil}$.
From our computations we found that $N_\mathrm{sf}$ varies as the maximum value of $P$ (data not shown).
No equivalent quantities can be identified that are intrinsic to the focal adhesion, as it always remains far from equilibrium and, consequently, its length $\hat{x}_\mathrm{fa}$, and centroid position $\tilde{x}_\mathrm{fa}$, are always changing.

The time to failure is a relevant quantity for systems collapsing due to full resorption of the focal adhesion.
A contour plot of this parameter is shown in Fig. \ref{fig:contours-unstretched} for configurations in region \textit{FA-c} of the response map.
The time to failure rapidly decreases for decreasing $\hat{x}_\mathrm{fa}^0$, and slowly decreases for increasing $x_\mathrm{sf}^0$, i.e for configurations far from the boundary between regions \textit{R} and \textit{FA-c}.

For large values of $\hat{x}_\mathrm{fa}^0$ (region \textit{SF-c} in the response map of Fig. \ref{fig:phase-d-un}) the system always collapses due to complete disassembly of the stress fiber over very short time scales (dynamic data not shown); a large focal adhesion acts as a very stiff support, allowing the force within the system, and hence the strain energy, to increase and drive the stress fiber to a rapid disassembly.

\section*{Collapse-mechanisms and system behavior with applied strain}
\label{sect:maps-stretch}
\begin{figure}[t]
\centering
\includegraphics{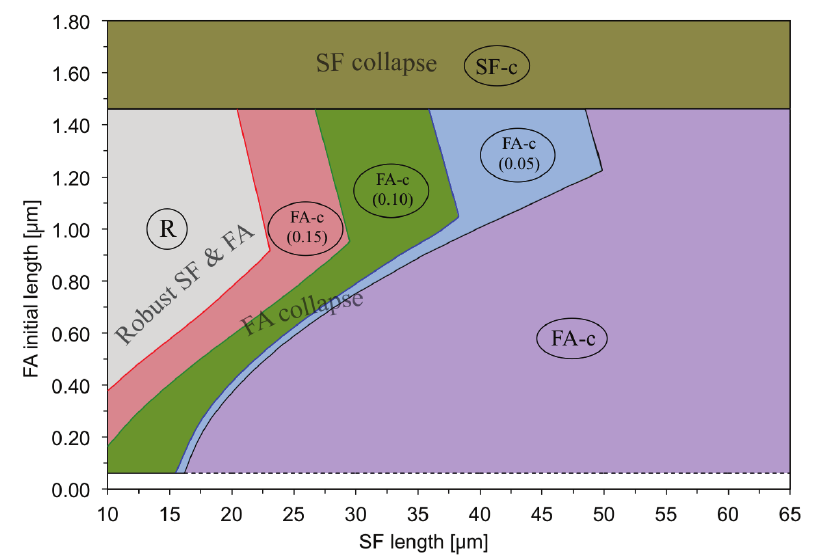}
\caption{\small{System response map for applied strain. Regions \textit{R} and \textit{FA-c} are modified from Fig. \ref{fig:phase-d-un}. The applied strain appears in parentheses in subregions of \textit{FA-c}.}}
\label{fig:phase-d-str}
\end{figure}
Fig. \ref{fig:phase-d-str} depicts the system response map under an applied step strain.
The numbers in parentheses are the strains for which failure occurs by focal adhesion resorption for that configuration in region \textit{FA-c} (see the discussion on Fig. \ref{fig:stretched-evos}, and Fig. \ref{fig:stretched-evos-pillar}).
On comparing with the response map under no strain in Fig. \ref{fig:phase-d-un}, it is apparent that region \textit{FA-c} has grown at the expense of \textit{R}; this suggests that, upon stretching, the system is more prone to collapse due to focal adhesion resorption.
The region in which the system does survive is restricted to initial configurations with progressively smaller $x_\mathrm{sf}^0$ and larger $\hat{x}_\mathrm{fa}^0$.

Our model admits substrate strains that are arbitrary functions of time, but we chose to apply time-discontinuous strains to make connections with the results of Mann et al. \citep{mann:2012}.
The strain was always applied at $t = 1800 \ s$, well after the system had attained a near-equilibrium state characterized by $N_\mathrm{sf}$ and $\hat{x}_\mathrm{fa}$ being steady, and the contractile force in a near-plateau regime (Fig. \ref{fig:stretched-evos}).
As in the unstretched test-cases, $\hat{x}_\mathrm{fa}^0$ and $x_\mathrm{sf}^0$ were varied; additionally, time-discontinuous strains of $0.05$, $0.10$ and $0.15$ were applied to the system by varying the stretch of the underlying substrate.

\section*{Time-dependent response of the system under different levels of strain}
\label{sect:evolutions-stretched}
The analysis of the detailed dynamics of the system for different strain amplitudes allows a greater appreciation of the effects of an external strain to the system and enables a more direct comparison with the experiments conducted by Mann et al. \citep{mann:2012}, in which two different levels of stretch were applied to the cells.
\begin{figure*}[h]
\centering
\includegraphics{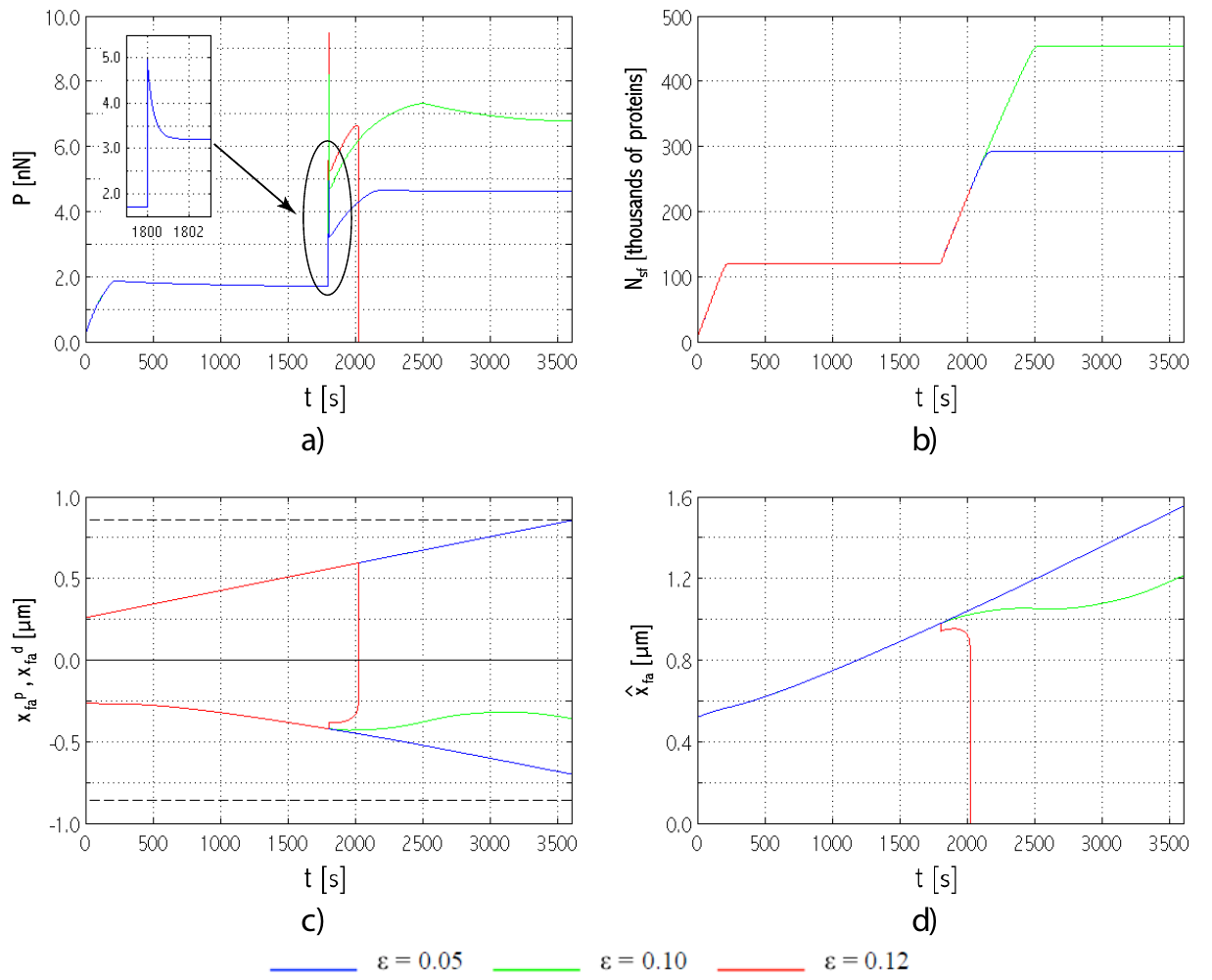}
\caption{\small{Time evolution of   a) force, $P$; b) number of actin monomers in the stress fiber, $N_\mathrm{sf}$; c) focal adhesion distal (initially negative values, $x^\mathrm{d}_\mathrm{fa}$) and proximal (positive values, $x^\mathrm{p}_\mathrm{fa}$) end positions; and d) focal adhesion length ($\hat{x}_\mathrm{fa}$) for the indicated applied strain. System initial configuration: $x_\mathrm{sf}^0 = 18 \mu m$; $\hat{x}_\mathrm{fa}^0 = 0.520 \mu m$. In c) the dashed black lines indicate the position of the micropost edges.}}
\label{fig:stretched-evos}
\end{figure*}
The plots in Fig. \ref{fig:stretched-evos} show the system dynamics for applied strains of $0.05, ~0.10$ and $0.12$ (the initial geometric configuration being fixed in order to allow a meaningful comparison between the different cases).
Upon stretching, $P$ spikes instantaneously (Fig. \ref{fig:stretched-evos}a) because of the elastic response of the system.
The force then drops very rapidly due to the passive viscoelastic response of the stress fiber.
The inset  in Fig. \ref{fig:stretched-evos}a shows these elastic and viscoelastic responses at a finer force-time resolution for the applied strain of $0.05$.
The externally applied strain also drives the dynamics of the stress fiber (Fig. \ref{fig:stretched-evos}b): more actin monomers are recruited, and the stress fiber grows until a second critical value of $N_\mathrm{sf}$ is reached.
As a consequence, $P$ rises again, driven by $P_\mathrm{sf}^\mathrm{ac}$, until it reaches a second maximum (this will be referred to as the global maximum force for that strain) followed by a second near-plateau, with a slightly negative slope.
Notably, the global maximum of $P$ and the post-strain critical value of $N_\mathrm{sf}$ increase if the applied strain increases.
An exception, however, occurs if the system experiences focal adhesion collapse: in Fig. \ref{fig:stretched-evos}a, for instance, the global maximum of $P$ for the strain of $0.12$ is lower than that for the strain of $0.10$.

The focal adhesion has a greater range of responses than the stress fiber (Fig. \ref{fig:stretched-evos}c and \ref{fig:stretched-evos}d). The proximal end always grows upon stretching, while the distal end can either suffer an initial resorption followed by restoration of the growth regime (green curve in Fig. \ref{fig:stretched-evos}c, strain of $0.10$) or grow monotonically (blue curve in Fig. \ref{fig:stretched-evos}c, strain of $0.05$).
Consequently, the focal adhesion can either have a transitory resorption stage or show monotonic growth (Fig. \ref{fig:stretched-evos}d).
In contrast to $P$, the focal adhesion length decreases for increasing strain (Fig. \ref{fig:stretched-evos}d, strain of $0.10$ \textit{versus} $0.05$).
At higher applied strains the focal adhesion begins to shrink irreversibly, causing the system to collapse (red curves in Fig. \ref{fig:stretched-evos}).
%
\section*{DISCUSSION}
\label{sect:disc}
The key to deciphering the system's complex mechano-chemical coupling lies with the chemical potentials of the stress fiber, focal adhesion and cytosol and with the complex, non-linear mechano-chemical coupling in the model.
On this basis, in the following subsections we highlight some aspects of the dynamics of the model that will be relevant to the discussion of the results presented in this paper.
\subsection*{Critical loads for assembly and disassembly}
\begin{figure}[h]
\centering
\includegraphics{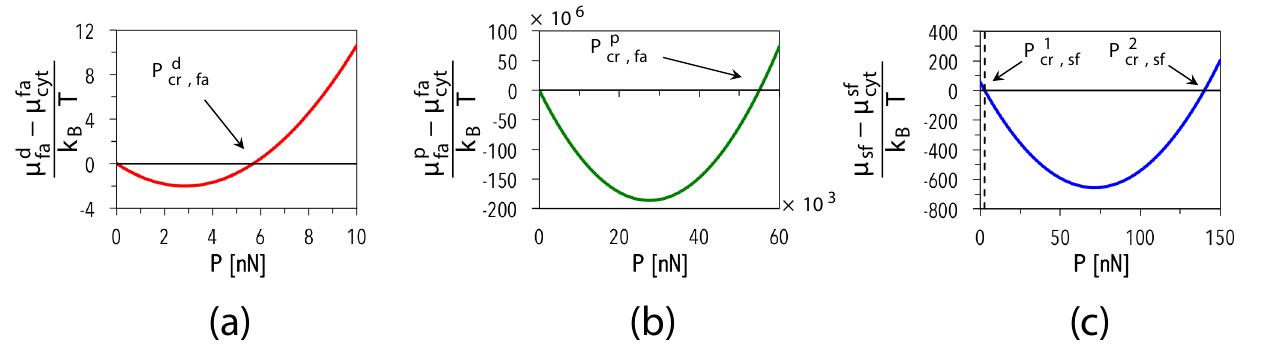}
\caption{\small{Chemical potentials as functions of the force $P$ at: a) the focal adhesion distal end; b) the focal adhesion proximal end; c) the stress fiber for the set of parameters listed in Tab. \ref{tab:params} in Appendix.}}
\label{fig:chem-pots}
\end{figure}

The chemical potentials that drive stress fiber and focal adhesion dynamics are themselves functions of the force, $P$, developed within the system (Fig. \ref{fig:chem-pots}).
By comparing $P$ with suitable critical values it can be established whether the focal adhesion or stress fiber undergoes growth or disassembly.
It is important to recognize, however, that these critical values vary, because they depend upon $N_\mathrm{sf}$ and $c_\mathrm{fa}$, which evolve.

With regard to the focal adhesion sub-system, experiments show that no growth is observed in the absence of force \citep{balaban:2001,riveline:2001}; for this reason, all the parameters were chosen such that $\mu_\mathrm{fa}-\mu^\mathrm{fa}_\mathrm{cyt} = 0$ if $P = 0$ (see Fig. \ref{fig:chem-pots}a and Fig. \ref{fig:chem-pots}b).
As a result, only one critical value of $P$ can be identified for both the distal and the proximal ends of the focal adhesion (namely $P_\mathrm{cr, fa}^\mathrm{d}$ and $P_\mathrm{cr, fa}^\mathrm{p}$ in Fig. \ref{fig:chem-pots}); below this force, the chemical potential drives focal adhesion complexes to bind, whereas above it unbinding is experienced at the given focal adhesion end.
For $P > P_\mathrm{cr, fa}^\mathrm{ d}$, it is the growth rate at the focal adhesion proximal end that determines whether the focal adhesion as a whole undergoes growth, translation, or resorption leading to eventual focal adhesion collapse; nevertheless, $P > P_\mathrm{cr, fa}^\mathrm{ d}$ is a necessary condition for focal adhesion resorption.

Given the parameter values chosen for the present study and system configurations explored, the critical load $P_\mathrm{cr, fa}^\mathrm{ p}$ -- above which $(\mu^\mathrm{p}_\mathrm{fa}-\mu^\mathrm{fa}_\mathrm{cyt})$ becomes positive leading to protein unbinding at the proximal end -- is much greater than the value of $P$ observed in our simulations.
Hence, it is not of interest.

On the other hand, for the set of parameters used here, two critical loads can be identified for the stress fiber sub-system (namely $P_\mathrm{cr, sf}^1$ and $P_\mathrm{cr, sf}^2$ in Fig. \ref{fig:chem-pots}c).
The dynamics of the sub-system are therefore dictated by comparing $P$ with such critical forces; in particular, for $P < P_\mathrm{cr, sf}^1$ or $P > P_\mathrm{cr, sf}^2$, the term $(\mu_\mathrm{sf} - \mu^\mathrm{sf}_\mathrm{cyt})$ is positive and the stress fiber undergoes disassembly, whereas for $P_\mathrm{cr, sf}^1 < P < P_\mathrm{cr, sf}^2$, the term $(\mu_\mathrm{sf} - \mu^\mathrm{sf}_\mathrm{cyt})$ is negative, and proteins are recruited to the stress fiber.

Our mechano-chemical model highlights the interplay between the mechanics and chemistry in determining the dynamics of the system.
Through the chemical potential, the force in the system affects the protein binding and unbinding rates, which determine the focal adhesion length and the stress fiber thickness.
In turn, these system geometric parameters influence the chemical potentials by changing the critical loads.
They also control the passive and active contributions to the stress fiber force, and, ultimately, the force in the system, by varying the system stiffness and the number of motor proteins in the stress fiber.

\subsection*{Non-linearities, mechano-chemistry and response maps}
The relevant critical loads $P_\mathrm{cr,sf}^1$, $P_\mathrm{cr,sf}^2$ and $P_\mathrm{cr,fa}^\mathrm{d}$ are non-linear functions of the geometry of the system.
The relations between the force $P$ and these critical loads dictate assembly or disassembly of a sub-system. The overall system dynamics that yield the response maps in Fig. \ref{fig:phase-d-un} and Fig. \ref{fig:phase-d-str} depend on the rate of change of the critical loads with respect to that of $P$.
In the next few sections we will observe some aspects of the behavior of the system which arise from this mechanism.
In summary, in our model the stress fiber can reach a critical concentration only because $P_\mathrm{cr,sf}^1$ increases faster than $P$ and, after some time, the stress fiber reaches a configuration for which protein binding ceases (see Fig. \ref{fig:critical-load-region-A} and the associated discussion).
Similarly, but with opposite results, Fig. \ref{fig:critical-load-region-B} shows that the focal adhesion collapses because $P_\mathrm{cr,fa}^\mathrm{d}$ increases faster than $P$ and, after some time, the focal adhesion is in a configuration for which unbinding starts and proceeds at a increasingly faster rate (see the discussion related to Fig. \ref{fig:critical-load-region-B} for further details).  
The rate at which $P$ and the critical loads change is driven by the model's coupled mechano-chemistry, and by non-linearities in the constitutive relations for chemical potentials, mechanical forces and rate laws. These are critical to the form of the response maps (Figs. \ref{fig:phase-d-un} and \ref{fig:phase-d-str}).

\section*{Stress fiber growth stops when the critical actin concentration is reached}
\label{sect:critical-conc}
\begin{figure}[h]
\centering
\includegraphics{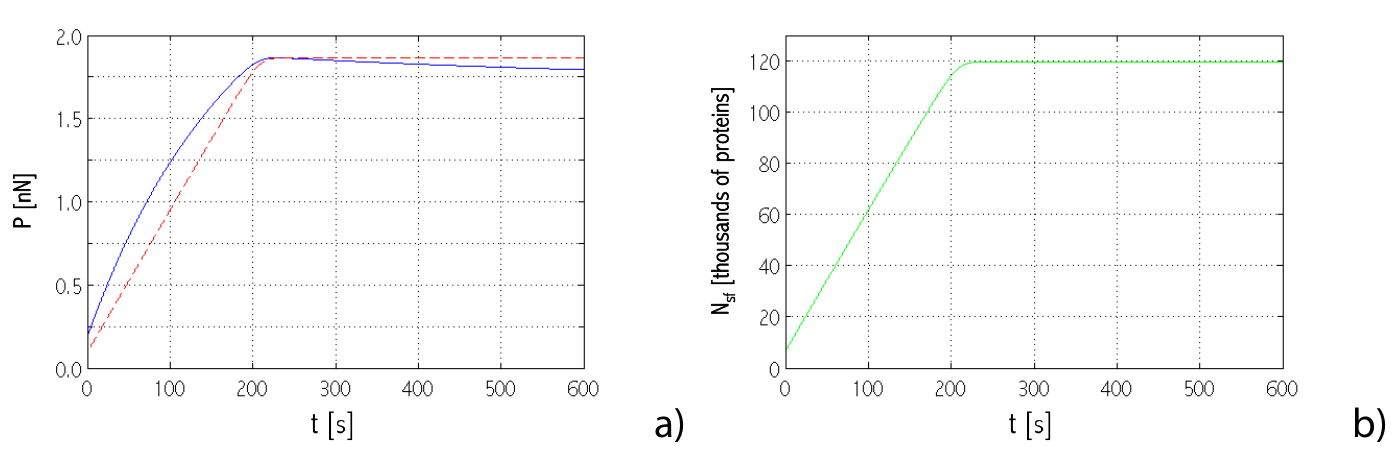}
\caption{\small{Time evolution of a) stress fiber force $P$ (blue line) and stress fiber critical force $P_\mathrm{cr,sf}^1$ (dashed red line); b) $N_\mathrm{sf}$. System initial configuration: $x_\mathrm{sf}^0 = 18 \mu \mathrm{m}$; $\hat{x}_\mathrm{fa}^0 = 0.520 \mu \mathrm{m}$ (region \textit{R} in Fig. \ref{fig:phase-d-un}).}}
\label{fig:critical-load-region-A}
\end{figure}

The attainment of a critical value of $N_\mathrm{sf}$ at which the stress fiber stops recruiting proteins, is explained by the evolution of $P$ relative to $P_\mathrm{cr,sf}^1$, as shown in Fig. \ref{fig:critical-load-region-A}a.
Initially, $P > P_\mathrm{cr, sf}^1$ makes $(\mu_\mathrm{sf} - \mu_\mathrm{sf}^\mathrm{cyt}) < 0$, which drives actin and myosin recruitment to the stress fiber (Fig. \ref{fig:critical-load-region-A}b).
Consequently, $P$ increases due to both enhanced acto-myosin contractility and the increased system mechanical stiffness (the stress fiber becomes thicker and the focal adhesions longer).
However, $P_\mathrm{cr, sf}^1$, which is a function of $N_\mathrm{act}$, also increases.
When $P_\mathrm{cr, sf}^1$ exceeds the stress fiber force, $(\mu_\mathrm{sf} - \mu^\mathrm{sf}_\mathrm{cyt}) >0$ and actin unbinding should occur.
However, $\chi_\mathrm{sf}$ in Eq. \ref{eq:sf-evo} is negative; therefore, actual unbinding rates remain low, and the stress fiber appears stable at its critical concentration (Fig. \ref{fig:critical-load-region-A}b).
Correspondingly, $P$ attains a near-plateau regime in which it slowly decreases under the effect of focal adhesion translation (see the discussion below on competition between stress fiber contractility and focal adhesion translation).

From this state, if $P$ increases due to \textit{external} perturbations to the system, but $P < P_\mathrm{cr,sf}^2$ is maintained, a growth regime can be re-established because the condition $(\mu_\mathrm{sf} - \mu^\mathrm{sf}_\mathrm{cyt}) < 0$ is regained.
Actin and myosin are then recruited until attainment of a second critical value of $N_\mathrm{sf}$ for which the stress fiber stops growing.
In the present study, the perturbation was applied in the form of a substrate strain (see Fig. \ref{fig:stretched-evos}).
A different perturbation induced by the finite cross-section of a micropost has been shown in Fig. \ref{fig:region-A-evos}, also.

\section*{Stress fiber activity can trigger different focal adhesion responses}
\label{sect:stress fiber-activity}
A longer stress fiber contains more myosin proteins and therefore is able to generate a higher active force, $P_\mathrm{sf}^\mathrm{ac}$.
For this reason, as shown in Fig. \ref{fig:contours-unstretched}, the maximum total force is higher for system configurations with longer stress fibers.
Secondarily, the passive contribution to the stress fiber force also plays a role in determining its maximum value. As Fig. \ref{fig:contours-unstretched} shows, focal adhesions that are initially large lead to systems developing higher forces, because the mechanical stiffness is higher.
Besides having a major effect on the the active force, stress fibers of different geometries also can trigger different focal adhesion responses: for instance, region \textit{R} in Fig. \ref{fig:phase-d-un} becomes increasingly narrow for longer stress fibers.
The reason is that in order to sustain the greater active force generated by a longer stress fiber, the initial focal adhesion needs to be longer. A longer focal adhesion has a higher critical load $P_\mathrm{cr, fa}^\mathrm{d}$ and can be subjected to a greater force without collapsing.
On the other hand, if the stress fiber is short, the active force generated is lower; hence, even focal adhesions developing from a single focal adhesion complex can sustain the load without failing (see the response map in Fig. \ref{fig:phase-d-un} and related discussion).

\begin{figure}[h]
\centering
\includegraphics{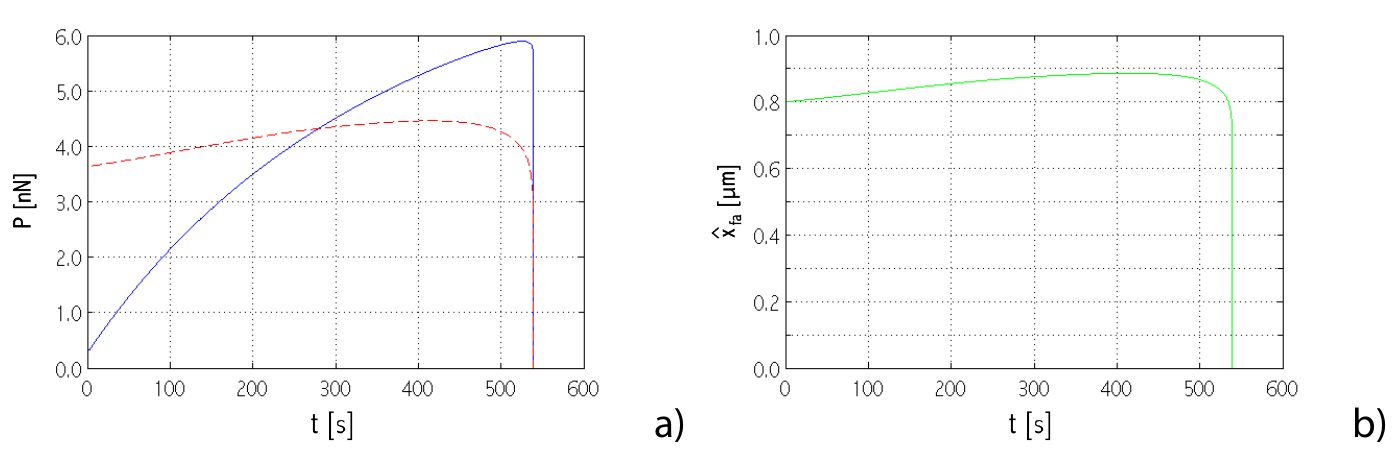}
\caption{\small{Time evolution of a) $P$ (blue line) and focal adhesion critical force, $P_\mathrm{cr, fa}^\mathrm{d}$ (dashed red line); b) $\hat{x}_\mathrm{fa}$. System initial configuration: $x_\mathrm{sf}^0 = 36 \mu \mathrm{m}$; $\hat{x}_\mathrm{fa}^0 = 0.800 \mu \mathrm{m}$ (region \textit{FA-c} in Fig. \ref{fig:phase-d-un}).}}
\label{fig:critical-load-region-B}
\end{figure}
Fig. \ref{fig:critical-load-region-B}a shows the evolution of both the total force, $P$, and the focal adhesion critical load, $P_\mathrm{cr, fa}^\mathrm{d}$, for a system with initial configuration in region \textit{FA-c} of Fig. \ref{fig:phase-d-un}. Due to the incorporation of more actins and myosins in the stress fiber, $P$ increases and exceeds $P_\mathrm{cr,fa}^\mathrm{d}$.
Then, the focal adhesion's growth slows down (Fig. \ref{fig:critical-load-region-B}b) because unbinding occurs at the distal end (as a consequence, the focal adhesion critical load also increases more slowly).
However, as $N_\mathrm{sf}$ is far from its critical value, $P$ continues to increase above $P_\mathrm{cr,fa}^\mathrm{d}$, eventually leading to severe resorption at the distal end, and focal adhesion collapse (Fig. \ref{fig:critical-load-region-B}b). Protein resorption is boosted by the force-dependent term $\chi_{fa}$ in Eq. \ref{eq:fa-evo-d}, which makes the unbinding rate grow exponentially with the stress fiber force.

\section*{The focal adhesion size can determine the fate of the stress fiber}
\label{sect:focal adhesion-activity}
For initial configurations in region \textit{SF-c} of Fig. \ref{fig:phase-d-un}, $\hat{x}_\mathrm{fa}^0$ is large and the stress fiber disassembles within the first few milliseconds of the computation.
The reason is that a large $\hat{x}_\mathrm{fa}^0$ makes the stress fiber-focal adhesion system mechanically very stiff.
Therefore, contractility drives $P$ to rapidly exceed $P_\mathrm{cr,sf}^2$, causing stress fiber disassembly.
The disassembly is boosted by the force-dependent term $\chi_{sf}$ in Eq. \ref{eq:sf-evo}, which enhances the actin unbinding rate.
The focal adhesion thus can control the fate of the system, by acting as a very stiff support.
\section*{Competition between stress fiber contractility and focal adhesion translation determines the force behavior}
\label{sect:treadmilling-vs-contractility}
For the system configurations in region \textit{R} of Fig. \ref{fig:phase-d-un} (or of Fig. \ref{fig:phase-d-str} for the applied strain case), the force reaches a plateau after an initial growth stage.
The slope of the plateau is regulated by the competition between stress fiber contractility and focal adhesion translation due to protein treadmilling.
The action of motor proteins in the stress fiber causes the active component of the stress fiber force $P_\mathrm{sf}^\mathrm{ac}$ (and, consequently, the total force $P$) to increase, whereas when the focal adhesion centroid moves towards the stress fiber, $P$ relaxes.
Our computations show that for the overall system dynamics this \textit{kinematic} relaxation mechanism and its interplay with stress fiber contractility is more relevant than the relaxation induced by passive viscoelasticity, because the latter occurs over very short time scales (see the inset in Fig. \ref{fig:stretched-evos}$a$).
For instance, for system configurations in region \textit{R}$^\prime$ of Fig. \ref{fig:phase-d-un}, the relaxation induced by focal adhesion translation towards the stress fiber has a major influence and prevails over the stiffening effect provided by the addition of myosin to the stress fiber.
Focal adhesion translation is enhanced for small values of $x_\mathrm{sf}^0$: the low values of the stress fiber force developed within the system lead to a large difference between the chemical potentials at the focal adhesion distal and proximal ends (Fig. \ref{fig:chem-pots}); thus, the binding rates of the focal adhesion ends prove to be very different.
This results in a high rate of focal adhesion translation, which in turn causes the stress fiber force to relax and vanish in a short time (blue curves in Fig. \ref{fig:region-A-evos}).

\section*{The influence of substrate loading on the overall system response}
\label{sect:disc-loading}
As shown in Fig. \ref{fig:stretched-evos}, larger external strains result in system collapse due to complete resorption of the focal adhesion.
A high strain leads to a high value of the force in the system, $P$, which can exceed the focal adhesion critical load, $P_{cr,sf}^\mathrm{d}$, and induce severe resorption at the focal adhesion distal end (as shown by the red curves in Fig. \ref{fig:stretched-evos}).

On the other hand, if the strain is sufficiently small, the stress fiber force reaches a second plateau; correspondingly, the stress fiber recruits more actins and myosins.
The focal adhesion also grows, demonstrating that to some extent an external load can stimulate growth of the stress fiber-focal adhesion system.
The strain is externally imposed as a substrate strain in our model, but in living cells may come from the ECM, neighboring cells or other stress fiber-focal adhesion complexes within the same cell.

\section*{Connection to recent cell traction force experiments on micropost arrays}
\begin{figure}[h]
\centering
\includegraphics{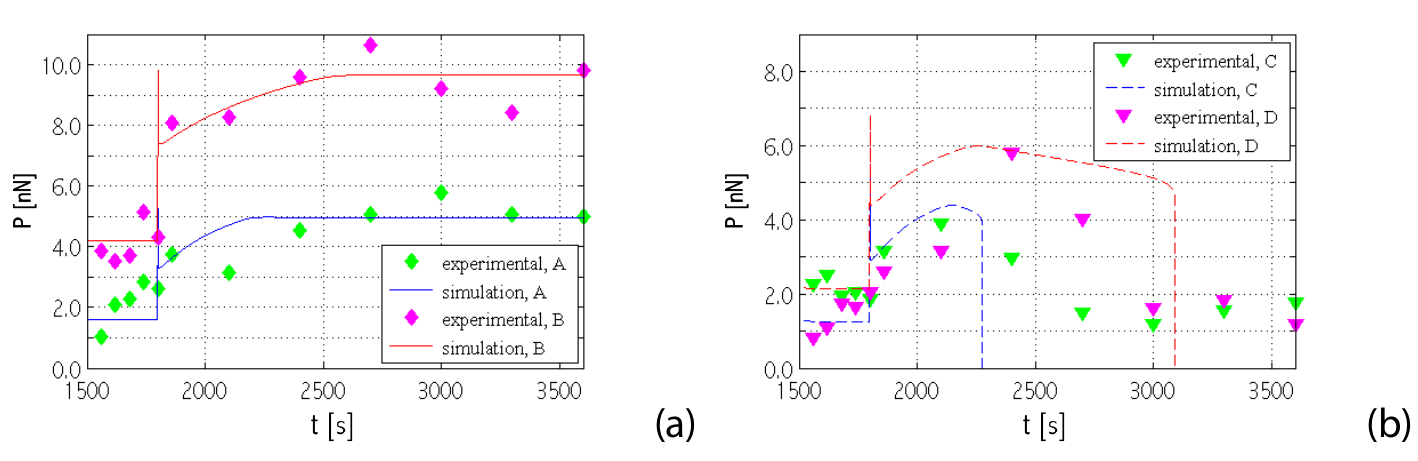}
\caption{\small{The computed stress fiber force \textit{versus} time compared with force on individual microposts from the work of Mann and co-workers \citep{mann:2012}. (a) Robust stress fiber-focal adhesion systems (Region R in Fig. \ref{fig:phase-d-str}). (b) Systems that suffer focal adhesion collapse (Region FA-c in Fig. \ref{fig:phase-d-str}). The strain of $0.06$ is applied at $1800$ s in both cases.}}
\label{fig:sim-vs-exp}
\end{figure}

Our results can be related to the experiments of Mann et al. on the force response of smooth muscle cells on arrays of polymeric microposts \citep{mann:2012}.
Fig. \ref{fig:sim-vs-exp} shows data from their study for the force on individual microposts \textit{versus} time in response to a substrate strain of $0.06$.
Fig. \ref{fig:sim-vs-exp}a corresponds to stress fiber-focal adhesion systems that remain robust over the period of the experiments -- Region R in Fig. \ref{fig:phase-d-str}.
Notably, the computed response has a spike in force at the instant of strain application due to the intrinsic viscoelastic response of the stress fiber, which has a relaxation time $\tau = 10$ s (see the inset in Fig. \ref{fig:stretched-evos}a and the related discussion, and Tab. \ref{tab:params} in Appendix).
The $1$-minute time resolution of the experiments was too coarse to capture such a spike.

In Fig. \ref{fig:sim-vs-exp}b stress fiber-focal adhesion systems from Region FA-c (focal adhesion collapse) of Fig. \ref{fig:phase-d-str} have been compared with experimental curves that show a significant decrease in force.
Notably, while the computed curves demonstrate decreases down to zero force, the experiments show a less sharp decreases followed by a plateau.
Upon examining the experimental force data we have found that the force trace on each of the two microposts represented in Fig. \ref{fig:sim-vs-exp}b is not complemented by a force trace that is equal in magnitude and opposite in direction on another micropost.
This suggests that while each end of a stress fiber is indeed connected to a focal adhesion on a micropost, different parts of the focal adhesion on these microposts have different stress fibers connected to them.
Each stress fiber and the part of the focal adhesions connected to each of its ends would form a system of the type considered in the model, and this system would have well-defined dynamics.
However, the force trace on a micropost is the magnitude of the vector resultant of all these different systems, some of which may collapse and all of which have different dynamics.
This yields the experimental curves in Fig. \ref{fig:sim-vs-exp}b characterized by sharply decreasing, but non-vanishing forces. In all cases, matches to the experimental curves were obtained by varying the initial focal adhesion length and the stress fiber unstretched length.

Mann et al. speculate that all the different observed behaviors may be due to the force acting on the focal adhesion before the application of the stretch.
Our study shows that the force does affect the system behavior,but is itself determined by the system's initial geometrical configuration -- Figs. \ref{fig:phase-d-un} and \ref{fig:phase-d-str}.
This diversity of stress fiber and focal adhesion geometries has not been reported by Mann et al.

\section*{Further capabilities of the model}
The discussion of Fig. \ref{fig:region-A-evos} identified an equilibrium state for the system when the focal adhesion grows to cover the micropost cross-section.
A non-uniform force distribution over the focal adhesion \citep{olberding:2010} also allows the attainment of an equilibrium state, but has not been considered here.

The model discussed here can be embedded in a whole cell model, where the effects of location within the cell and history, as well as of cell type, can be considered.
Notably, a) both stress fibers and focal adhesions vary in size and length throughout a cell, depending also on cell history, and from one cell type to another; b) the external strain field to which cells are subjected is non-uniform; and c) the kinetic rates of proteins binding/unbinding and the structural and chemical properties of both the stress fibers and the focal adhesions change with the cell type.
All these varying conditions, and the different responses they elicit, can be accounted for in the model presented here.

\section*{Acknowledgements}
We thank Prof. Jianping Fu for discussions, and for the use of experimental data.

\bibliographystyle{plainnat}      
\bibliography{biblio}
\clearpage
\appendix
\appendixpage
\section{Additional details on the model}
\label{sect:more-model}
\subsection{Constitutive relations for the chemical potentials and the mechanical forces}
The expressions for the chemical potentials are:
\begin{subequations} \label{eq:chem-pots}
\begin{align}
\mu^\mathrm{sf}_\mathrm{cyt} &= H_\mathrm{cyt}^\mathrm{sf} + k_B T \ \mathrm{ln}\bracs{\widehat{N}_\mathrm{sf}/ \brac{N_\mathrm{sf}^\mathrm{max} - \widehat{N}_\mathrm{sf}} } \label{eq:chem-pots_cyt_sf} \\
\mu_\mathrm{sf} &= \frac{1}{2} \frac{\brac{P x^0_\mathrm{sf}}^2}{E_\mathrm{sf} N_\mathrm{sf}^2 V_\mathrm{act}} + U_\mathrm{sf}^\mathrm{conf} - \frac{P}{N_\mathrm{fil}} \ d_\mathrm{sf}. \label{eq:chem-pots_sf} \\
\mu^\mathrm{fa}_\mathrm{cyt} &= H_\mathrm{cyt}^\mathrm{fa} + k_B T \ \mathrm{ln}\bracs{\widehat{N}_\mathrm{fa}/ \brac{N_\mathrm{fa}^\mathrm{max} - \widehat{N}_\mathrm{fa}} } \label{eq:chem-pots_cyt_fa}\\
\mu_\mathrm{fa}^\mathrm{d} &= \frac{1}{2} \frac{P^2 h}{\overline{E}_\mathrm{fa} c_\mathrm{fa}^\mathrm{max} \hat{x}^2_\mathrm{fa} b} \!+\! \frac{1}{2} B \kappa^2 \lambda + U_\mathrm{fa}^{conf} \!-\! P \!\brac{\!d_\mathrm{fa} \!+\! \frac{\lambda}{2}\!} \label{eq:chem-pots_fa_d} \\
\mu_\mathrm{fa}^\mathrm{p} &= \frac{1}{2} \frac{P^2 h}{\overline{E}_\mathrm{fa} c_\mathrm{fa}^\mathrm{max} \hat{x}^2_\mathrm{fa} b} \!+\! \frac{1}{2} B \kappa^2 \lambda + U_\mathrm{fa}^{conf} \!-\! P \!\brac{\!d_\mathrm{fa} \!-\! \frac{\lambda}{2}\!}. \label{eq:chem-pots_fa_p}
\end{align}
\end{subequations}
The expressions for the constitutive equations that relate the force to the stress fiber stretch, the focal adhesion deformation and the micropost displacement are:
\begin{subequations} \label{eq:forces}
\begin{align}
P_\mathrm{sf} &= 
                 \pi r_\mathrm{sf}^2 E_\mathrm{sf} \bracs{\gamma_\mathrm{e} \brac{\frac{x_\mathrm{sf}}{x_\mathrm{sf}^0} - 1}
               + \gamma_\mathrm{ve} \int\limits_{0}^{t} \frac{\dot{x}_\mathrm{sf}\brac{s}}{x_\mathrm{sf}^0} e^{- \brac{t - s}/\tau} \, ds}
               + \underbrace{\frac{P_\mathrm{sf}^\mathrm{stl}}{\dot{\varepsilon}^\mathrm{con}} \brac{\dot{\varepsilon}^\mathrm{con} - \frac{\dot{x}_\mathrm{sf}}{x_\mathrm{sf}^0}} }_{P_\mathrm{sf}^\mathrm{ac}}
\label{eq:force-sf} \\
P_\mathrm{fa} &= \frac{ \overline{E}_\mathrm{fa} \hat{x}_\mathrm{fa} b}{h} x_\mathrm{fa}^\mathrm{e}& \label{eq:force-fa} \\
P_\mathrm{mp} &= \frac{3 \pi E_\mathrm{mp} r_\mathrm{mp}^4}{4 h_\mathrm{mp}^3} x_\mathrm{mp}& \label{eq:force-mp}
\end{align}
\end{subequations}
An explanation of all the parameters appearing in Eq. \eqref{eq:chem-pots} and Eq. \eqref{eq:forces} can be found in Tab. \ref{tab:params} and in Maraldi and Garikipati \citep{maraldi:2013}.
We note that the term in square brackets in Eq. (\ref{eq:force-sf}) is the standard linear solid viscoelastic model, chosen because it allows full invertibility between force and displacement responses \citep{maraldi:2013}. 
The stress fiber's active contractile force $P_\mathrm{sf}^\mathrm{ac}$, given by the last term in Eq. (\ref{eq:force-sf}), results from the acto--myosin contractile units introduced in Fig. \ref{fig:model} and discussed there.
As also explained in the Main Text, the number of acto--myosin contractile units is proportional to $N_\mathrm{sf}$, and each unit in the stress fiber is taken to have the same strain rate $\dot{x}_\mathrm{am}/x_\mathrm{am}^0$, where $x_\mathrm{am}$ is the deformed (contracted) length, and $x_\mathrm{am}^0$ the reference length of the unit.
The maximum contractile speed of a myosin motor, $\dot{x}^\mathrm{con}_\mathrm{myos}$ gives the maximum contractile strain rate of the stress fiber: $-\dot{x}^\mathrm{con}_\mathrm{myos}/x_\mathrm{sf}^0$, which we denote as $\dot{\varepsilon}^\mathrm{con}$ for brevity.
This is a constant in our model.
The contractile force generated by a contractile unit is given by
\begin{equation}
P_\mathrm{am}^\mathrm{ac} = \frac{P_\mathrm{am}^\mathrm{stl}}{\dot{\varepsilon}^\mathrm{con}} \brac{\dot{\varepsilon}^\mathrm{con} - \frac{\dot{x}_\mathrm{am}}{x_\mathrm{am}^0}} 
\label{myosin-force}
\end{equation}
The quantity $P_\mathrm{am}^\mathrm{ac}$ acts as a force per contractile unit length along each actin filament that makes up the stress fiber.
The total contractile force in a single actin filament therefore is the sum of the force contributions from the contractile units along its length, and the total contractile force in the stress fiber is the sum of contractile forces in the parallel actin filaments that are bundled together to form the stress fiber.
The total contractile force in the stress fiber therefore is $P_\mathrm{sf}^\mathrm{ac} = P_\mathrm{am}^\mathrm{ac}\,\beta N_\mathrm{sf}$.
Here, $\beta$ is a constant of proportionality.
Since all actomyosin units in a stress fiber are taken to have the same strain rate, we can equivalently write $\dot{x}_\mathrm{sf}/x_\mathrm{sf}^0 = \dot{x}_\mathrm{am}/x_\mathrm{am}^0$. 
Finally, the additivity of contractile force from the units can be used to define $P_\mathrm{sf}^\mathrm{stl} = P_\mathrm{am}^\mathrm{stl}\,\beta N_\mathrm{sf}$.
These substitutions together with Eq. (\ref{myosin-force}) allow us to write
\begin{equation}
P_\mathrm{sf}^\mathrm{ac} = \frac{P_\mathrm{sf}^\mathrm{stl}}{\dot{\varepsilon}^\mathrm{con}} \brac{\dot{\varepsilon}^\mathrm{con} - \frac{\dot{x}_\mathrm{sf}}{x_\mathrm{sf}^0}} 
\label{sf-active-force}
\end{equation}
as used in Eq. (\ref{eq:force-sf}).

\subsection{Solution strategy}
For a given set of stress fiber stretch $x_\mathrm{sf}(t)$, focal adhesion length, centroid position and deformation -- $\hat{x}_\mathrm{fa}(t)$, $\tilde{x}_\mathrm{fa}(t)$ and $x^\mathrm{e}_\mathrm{fa}(t)$, respectively--, micropost displacement $x_\mathrm{mp}(t)$ and externally-applied substrate displacement $x_\mathrm{lyr}(t)$ at time $t$, where the time history $x_\mathrm{sf}(s)$ is known $\forall s \leq t$, the force within the system $P$ can be evaluated by assuming mechanical equilibrium \citep{maraldi:2013}.
The determination of $P$ is essential for calculating the chemical potentials of the focal adhesion, the stress fiber and the cytosol, which are the driving forces for the chemical processes \citep{maraldi:2013} and appear in the rate equations Eqs. (\ref{eq:sf-evo}--\ref{eq:fa-evo-p}).

The system of nonlinear ordinary differential equations Eqs.(\ref{eq:sf-evo}--\ref{eq:fa-evo-p}) is solved by semi-implicit time integration, using a linear-in-time approximation to evaluate the hereditary integral appearing in Eq.\eqref{eq:force-sf}.
As initial conditions, we specified the initial focal adhesion length $\hat{x}_\mathrm{fa}^0$ and the unstretched stress fiber length $x_\mathrm{sf}^0$.
Assuming that the focal adhesions are initially in the middle of the microposts allows the focal adhesion ends' initial positions $x^0_p$ and $x^0_d$ to be evaluated.
Furthermore, we assumed that the stress fiber initially consists of a single filament; this models the minimal precursor system.
Under this assumption and given the size of an actin monomer, the initial number of proteins in the stress fiber $N_\mathrm{sf}^0$ can be evaluated.

\clearpage
\section{Time-dependent response of the system in region \textit{FA-c}}
\label{sect:evolutions-unstretched-FAc}
\begin{figure}[h!]
\centering
\includegraphics{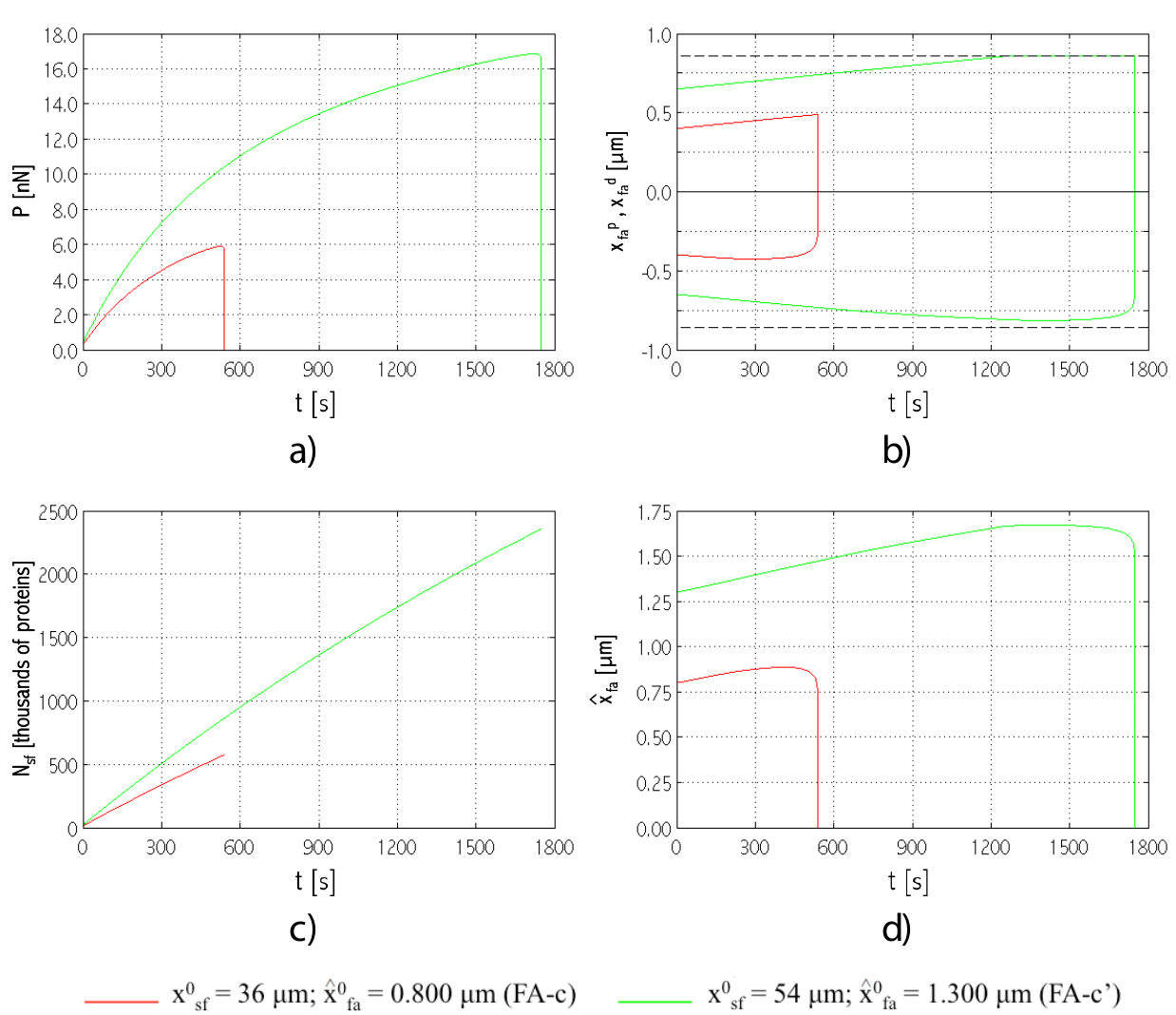}
\caption{\small{Time evolution of a) force, $P$; b) focal adhesion distal (initially negative values, $x^\mathrm{d}_\mathrm{fa}$) and proximal (positive values, $x^\mathrm{p}_\mathrm{fa}$) ends' positions; c) number of actin monomers in the stress fiber, $N_\mathrm{sf}$; and d) focal adhesion length, $\hat{x}_\mathrm{fa}$, for two different system initial configurations belonging to region \textit{FA-c} and \textit{FA-c'} in Fig.\ref{fig:phase-d-un}. In b) the dashed black lines indicate the position of the micropost edges.}}
\label{fig:region-BCD-evos}
\end{figure}

The red and the green curves in Fig.\ref{fig:region-BCD-evos} show the system's dynamics for configurations falling in region \textit{FA-c} and \textit{FA-c}$^\prime$ on the response map in Fig.\ref{fig:phase-d-un}, respectively.
Fig.\ref{fig:region-BCD-evos}b shows that both configurations suffer distal end unbinding of the focal adhesion, whereas the proximal end may either continue growing (red curve), or become stationary because it reaches the edge of the micropost (green curve).
In the latter case, overall focal adhesion growth slows down (Fig.\ref{fig:region-BCD-evos}$d$), as new protein complexes can only be added at the distal end.

As $P$ increases (Fig.\ref{fig:region-BCD-evos}a), the term $(\mu^\mathrm{d}_\mathrm{fa} - \mu^\mathrm{fa}_\mathrm{cyt})$ increases, causing the distal end binding rate to decrease; eventually, $(\mu^\mathrm{d}_\mathrm{fa} - \mu^\mathrm{fa}_\mathrm{cyt})$ becomes positive, unbinding starts and proceeds at an increasingly higher rate (boosted by the term $\chi_\mathrm{fa}$, Eq.\ref{eq:fa-evo-d}).
Then, the focal adhesion begins to shrink rapidly, until it is catastrophically resorbed (Fig.\ref{fig:region-BCD-evos}d).
The focal adhesion shrinkage makes $P$ decrease rapidly and, of course, when the focal adhesion is fully resorbed $P$ also vanishes.
For the chosen parameter values contributing to the chemical potentials $\mu_\mathrm{sf}$ and $\mu_\mathrm{cyt}^\mathrm{sf}$ in these cases, the stress fiber remains in a growth regime with continual recruitment of proteins (Fig.\ref{fig:region-BCD-evos}c) until the system itself collapses.

\clearpage
\section{More details on the dynamics of the stress fiber-focal adhesion system under applied external strain}
The response map resulting from our computations and reported in Fig.\ref{fig:phase-d-str} is obtained when the strain is applied to the system at $t = 1800$ s.
The extent of regions \textit{FA-c(0.05)}, \textit{FA-c(0.10)} and \textit{FA-c(0.15)} depicted in the map depends upon the time at which the step strain is applied.
In our model, in fact, the focal adhesion remains far from equilibrium and if the strain is applied at a later instant in time than that chosen here, it will have grown larger and will be able to sustain greater loads, because its critical force will be greater.
As a result, the strain necessary to cause system failure will be greater.
\begin{figure}[h!]
\centering
\includegraphics{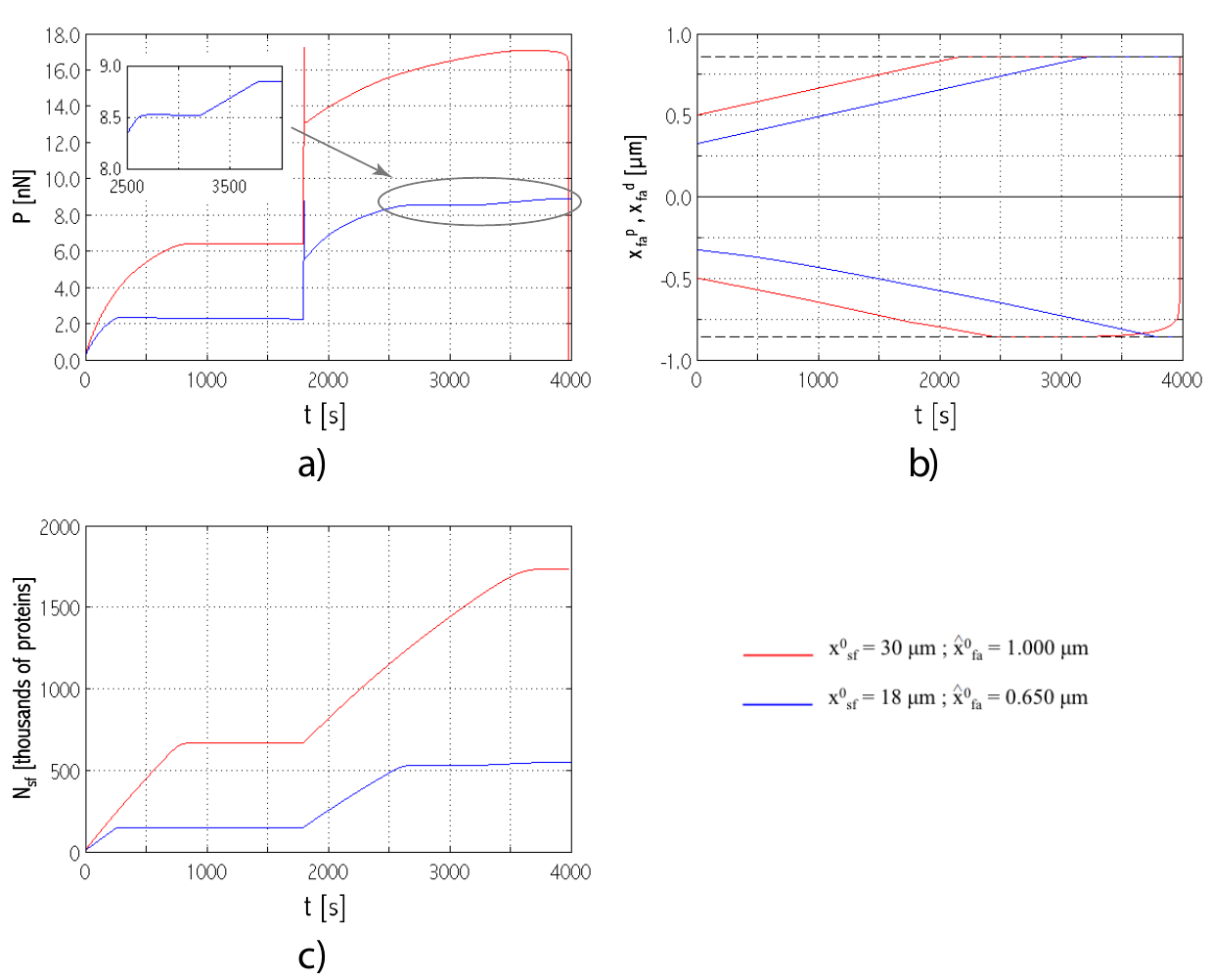}
\caption{\small{Time evolution of a) force, $P$; b) focal adhesion distal (initially negative values, $x^\mathrm{d}_\mathrm{fa}$) and proximal (positive values, $x^\mathrm{p}_\mathrm{fa}$) ends' positions; and c) number of actin monomers in the stress fiber, $N_\mathrm{sf}$, for two different system initial configurations and a step strain $\varepsilon = 0.1$. In b) the dashed black lines indicate the position of the micropost edges.}}
\label{fig:stretched-evos-pillar}
\end{figure}

An equilibrium configuration for the focal adhesion may be attained in some cases if both its end reach the edges of the micropost (blue curves in Fig.\ref{fig:stretched-evos-pillar}), as discussed in the section pertaining to the time-dependent response of the system with no applied strain in the Main Text.
Besides inducing a stiffening effect related to focal adhesion translation away from the stress fiber (as explained in the previous sections and shown in the blue curves in Fig.\ref{fig:stretched-evos-pillar}), when an external strain is applied to the system and for certain system initial configurations, the presence of a finite cross-section micropost may also cause a more rapid collapse of the system; this case is depicted in Fig.\ref{fig:stretched-evos-pillar} (red curves).
In both cases, the focal adhesion has grown to cover the cross-section of the micropost; however, if a focal adhesion length equal to the diameter of the micropost is not sufficient to sustain the force within the system (red curves), a rapid failure occurs.

\clearpage
\section{Parameters' values}
\vspace{-0.5cm}
\begin{table} [h]
\scriptsize
\centering
\rowcolors{1}{}{lightgray}
\begin{tabular} {|>{\scriptsize}l r l l >{\scriptsize}l|}
\hline
\bf{Parameter} & \bf{Symbol} & \bf{Value} & \bf{Unit} & \bf{Remarks} \\
\hline
\hline
FA effective elastic modulus & $\overline{E}_\mathrm{fa}$ & $5.5 \cdot 10^{6}$ & Pa & \parbox{6.5cm}{$= E_\mathrm{fa}$, elastic modulus assumed to be within the \\ percolation limit\\ \citep{maraldi:2013,olberding:2010}. \\ $E_\mathrm{fa}$ estim. for soft, gel-like biological materials} \\
FA width & $b$ & $5.0 \cdot 10^{-7}$ & m & \parbox{6.5cm}{Estimate from images in \cite{balaban:2001} and  \\ \cite{riveline:2001}} \\
FA height & $h$ & $1.0 \cdot 10^{-7}$ & m & \parbox{6.3cm}{Rough estimate on the basis of the length of some focal\\ adhesion proteins \citep{zamir:2001}} \\
Maximum attainable concentration & $c_\mathrm{fa}^\mathrm{max}$ & $1.72 \cdot 10^{7}$ & $\mathrm{m^{-1}}$ & $= 1 / \lambda$ \\
Binding enthalpy & $H^\mathrm{fa}_\mathrm{cyt}$ & $0.0$ & $\mathrm{N \cdot m^2}$ & \parbox{6.5cm}{Imposed, focal adhesion does not grow in absence of \\ force \citep{olberding:2010,balaban:2001}\\ \citep{riveline:2001}} \\
Cell membrane curvature & $\kappa$ & $4.0 \cdot 10^{5}$ & $\mathrm{m^{-1}}$ & Estimate from cell height $\sim 5 \mu\mathrm{m}$.\\
FA complex length & $\lambda$ & $5.8 \cdot 10^{-8}$ & m & From ref. \citep{arnold:2004} \\
\parbox{3.7cm}{Change in internal energy due to \\binding conformational changes} & $U_\mathrm{fa}^\mathrm{conf}$ & $0.0$ & J & \parbox{6.5cm}{Imposed, focal adhesion does not grow in absence of \\ force \citep{olberding:2010,balaban:2001}\\ \citep{riveline:2001}} \\
\parbox{3.7cm}{Equivalent displacement due to \\binding conformational changes} & $d_\mathrm{fa}$ & $2.9006 \cdot 10^{-8}$ & m & \parbox{6.5cm}{Set to reproduce a variety of observed experimental\\ behaviors} \\
\parbox{3.5cm}{Chemical potential of FA \\proteins in the cytosol} & $\mu^\mathrm{fa}_\mathrm{cyt}$ & $0.0$ & J & \parbox{6.5cm}{Imposed, focal adhesion does not grow in absence of \\force \citep{olberding:2010,balaban:2001} \\ \citep{riveline:2001}} \\
FA proteins binding rate & $k_\mathrm{fa}^\mathrm{b}$ & $2.85 \cdot 10^{-3}$ & $\mathrm{s^{-1}}$ & \parbox{6.5cm}{Set to reproduce a variety of observed experimental \\behaviors} \\
FA proteins unbinding rate & $k_\mathrm{fa}^\mathrm{u}$ & $7.98 \cdot 10^{-4}$ & $\mathrm{s^{-1}}$ & \parbox{6.5cm}{Set to reproduce a variety of observed experimental \\behaviors} \\
SF Young's modulus & $E_\mathrm{sf}$ & $8.0 \cdot 10^{7}$ & Pa & Estimate based on \cite{deguchi:2006} \\
Actin monomer volume & $V_\mathrm{act}$ & $1.047 \cdot 10^{-25}$ & $\mathrm{m^3}$ & \parbox{6.5cm}{Calculated from data on actin length and diameter in \\ \cite{howard:2001}} \\
\parbox{3.7cm}{Equivalent displacement due to \\binding conformational changes} & $d_\mathrm{sf}$ & $2.32 \cdot 10^{-9}$ & m & \parbox{6.5cm}{Set to reproduce a variety of observed experimental \\behaviors} \\
Actin monomer length & $L_\mathrm{actmon}$ & $2.72 \cdot 10^{-9}$ & m & From \cite{howard:2001} \\
\parbox{3.7cm}{Change in internal energy due to \\binding conformational changes} & $U_\mathrm{sf}^\mathrm{conf}$ & $0.0$ & J & Absorbed into $H_\mathrm{cyt}^\mathrm{sf}$ \\
Binding enthalpy & $H_\mathrm{cyt}^\mathrm{sf}$ & $-2.47 \cdot 10^{-19}$ & J & \parbox{6.5cm}{Set to reproduce a variety of observed experimental \\behaviors} \\
Maximum SF proteins concentration & $c_\mathrm{sf}^\mathrm{max}$ & $1.144 \cdot 10^{11}$ & $\mathrm{m^{-1}}$ & \parbox{6.5cm}{Estim. from total number of actin monomers in yeast \\cytosol \citep{wu:2005}, ratio of volume\\of cell to yeast cell and considering $50$ SFs with \\mean length of $10 \mu\mathrm{m}$ in a cell} \\
SF proteins binding rate & $k_\mathrm{sf}^\mathrm{b}$ & $2.725 \cdot 10^{-4}$ & $\mathrm{s^{-1}}$ & \parbox{6.5cm}{Adapted from association rate for ATP-actin at the \\barbed end \citep{pollard:2003}} \\
SF proteins unbinding rate & $k_\mathrm{sf}^\mathrm{u}$ & $0.8$ & $\mathrm{s^{-1}}$ & \parbox{6.5cm}{Adapted from association rate for ATP-actin at the \\pointed end \citep{pollard:2003}} \\
\parbox{3.5cm}{Non-dimensional modulus\\ (elastic branch)} & $\gamma_\mathrm{e}$ & $0.9$ & & \parbox{6.5cm}{Set to reproduce a variety of observed experimental \\behaviors} \\
\parbox{3.5cm}{Non-dimensional modulus\\ (viscoelastic branch)} & $\gamma_\mathrm{ve}$ & $0.1$ & & \parbox{6.5cm}{Set to reproduce a variety of observed experimental \\behaviors} \\
Relaxation time & $\tau$ & $10.0$ & s & Estimate from Fig. 3 in \cite{kumar:2006} \\
SF maximum contractile velocity & $\dot{x}_\mathrm{myos}^\mathrm{con}$ & $-5.0 \cdot 10^{-7}$ & $\mathrm{m \cdot s^{-1}}$ & From \cite{lord:2004} \\
Myosin stalling force & $P_\mathrm{myos}^\mathrm{stl}$ & $3.0 \cdot 10^{-11}$ & N & From \cite{wu:2005} \\
Myosin-to-actin proteins ratio & $\beta$ & $1.08 \cdot 10^{-3}$ & & From \cite{wu:2005} \\
Micropost Young's modulus & $E_\mathrm{mp}$ & $2.5 \cdot 10^{6}$ & Pa & From \cite{mann:2012} \\
Micropost radius & $r_\mathrm{mp}$ & $9.15 \cdot 10^{-7}$ & m & From \cite{mann:2012} \\
Micropost height & $h_\mathrm{mp}$ & $8.3 \cdot 10^{-6}$ & m & From \cite{mann:2012} \\
Boltzmann's constant & $k_\mathrm{B}$ & $1.381 \cdot 10^{-23}$ & $\mathrm{J \cdot K^{-1}}$ & \\
Test temperature & $T$ & $310.0$ & K & \\
\hline
\end{tabular}
\caption{\small{Parameters values used in the computations.}}
\label{tab:params}
\end{table}

\end{document}